\documentclass{elsart4-1}
\usepackage{amssymb}
\usepackage{graphics}
\usepackage[english,francais]{babel}

\newtheorem{e-proposition}[theorem]{Proposition}

\newtheorem{e-definition}[theorem]{Definition\rm}

\renewcommand{\theequation}{\arabic{equation}}
\def\og{\leavevmode\raise.3ex\hbox{$\scriptscriptstyle\langle\!\langle$~}}
\def\fg{\leavevmode\raise.3ex\hbox{~$\!\scriptscriptstyle\,\rangle\!\rangle$}}

\usepackage [dvips]{graphicx}
\usepackage{amsmath}
\usepackage{epsfig}
\usepackage{amssymb}

\newcommand{\cL}{\mathcal L}
\newcommand{\hcL}{{{\mathcal L}}}
\newcommand{\hr}{{\hat \rho}}

\newcommand{\bG}{{\Gamma}}
\newcommand{\dd}{\text{d}}

\newcommand{\ee}{\text{e}}

\newcommand{\p}{\partial}

\newcommand{\br}{\text{\bf r}}

\newcommand{\bv}{\text{\bf v}}

\newcommand{\bsigma}{\boldsymbol{\sigma}}
\newcommand{\bs}{\boldsymbol{\sigma}}
\newcommand{\bsig}{\boldsymbol{\sigma}}

\begin{document}

\centerline{Physics}
\begin{frontmatter}


\selectlanguage{english}
\title{Power injected in a granular gas}


\selectlanguage{english}
\author[paolo]{Paolo Visco},
\ead{Paolo.Visco@th.u-psud.fr}
\author[andrea]{Andrea Puglisi}
\ead{andrea.puglisi@roma1.infn.it}
\author[alain]{Alain Barrat}
\ead{Alain.Barrat@th.u-psud.fr}
\author[emmanuel]{Emmanuel Trizac}
\ead{Emmanuel.Trizac@lptms.u-psud.fr}
\author[paolo]{Fr\'ed\'eric van Wijland}
\ead{Frederic.van-Wijland@th.u-psud.fr}

\address[paolo]{Laboratoire Mati\`ere et Syst\`emes Complexes,
(CNRS UMR7057), Universit\'e de Paris VII -- Denis Diderot,
10 rue Alice Domon et L\'eonie Duquet, 75205 Paris cedex 13, France}
\address[andrea]{Dipartimento di Fisica, Universit\`a La Sapienza, Piazzale A. Moro 2, 00185, Rome, Italy}
\address[alain]{Laboratoire de Physique Th\'eorique (CNRS UMR8627), B\^atiment 210,
Universit\'e Paris-Sud, 91405 Orsay cedex, France}
\address[emmanuel]{Laboratoire de Physique Th\'eorique et Mod\`eles Statistiques
(CNRS UMR8626), B\^atiment 100, Universit\'e Paris-Sud, 91405 Orsay cedex, France}


\medskip
\begin{center}
{\small Received *****; accepted after revision +++++}
\end{center}

\begin{abstract}
  A granular gas may be modeled as a set of hard-spheres undergoing
  inelastic collisions; its microscopic dynamics is thus strongly
  irreversible. As pointed out in several experimental works bearing
  on turbulent flows or granular materials, the power injected in a
  dissipative system to sustain a steady-state over an asymptotically
  large time window is a central observable. We describe an analytic
  approach allowing us to determine the full distribution of the power
  injected in a granular gas within a steady-state resulting from
  subjecting each particle independently either to a random force
  (stochastic thermostat) or to a deterministic force proportional to
  its velocity (Gaussian thermostat). We provide an analysis of our
  results in the light of the relevance, for other types of systems,
  of the injected power to fluctuation relations.
 
\vskip 0.5\baselineskip

\selectlanguage{francais}
\noindent{\bf R\'esum\'e}
\vskip 0.5\baselineskip
\noindent
{\bf Puissance inject\'ee dans un gaz granulaire.}
Un gaz granulaire est assimil\'e \`a une assembl\'ee de sph\`eres
dures in\'elastiques, ce qui conf\`ere \`a sa dynamique microscopique
un caract\`ere fortement irr\'eversible. Ainsi que sugg\'er\'e par
plusieurs travaux exp\'erimentaux, r\'ealis\'es sur des fluides
turbulents ou granulaires, la puissance inject\'ee dans ces syst\`emes
pour les maintenir dans un \'etat stationnaire sur une grande fen\^etre
temporelle est une observable de choix. Nous pr\'esentons une
m\'ethode analytique permettant d'acc\'eder \`a la distribution de la
puissance inject\'ee dans un gaz granulaire au sein d'un \'etat
stationnaire obtenu en soumettant chaque particule soit \`a une force
al\'eatoire (thermostat stochastique), soit \`a une force
d\'eterministe proportionnelle \`a sa vitesse (thermostat gaussien).
Nous analysons nos r\'esultats \`a la lumi\`ere du r\^ole jou\'e, dans
d'autres types de syst\`emes, par la puissance inject\'ee dans le
contexte des relations de fluctuation.

\vskip 0.5\baselineskip
\noindent{\small{\it Mots-cl\'es~:} gaz granulaire~; grandes d\'eviations~;
\'etat stationnaire hors de l'\'equilibre}}
\end{abstract}
\end{frontmatter}

\selectlanguage{english}
\section{Motivations}
\label{motiv}
One of the lessons of thermodynamics is that in order to characterize
the macroscopic properties of systems in equilibrium only a few numbers
of well-chosen variables are necessary. Intensive state
functions, like the free energy per unit volume, depending on
intensive variables only (like temperature and density), can then be used
to determine {\it e.g.}  the system's phase diagram. Statistical
mechanics is the theory that precisely allows to bridge the
microscopics to the interesting macroscopic behavior, which it does by
bypassing the details of the dynamical rules governing the time
evolution of the system at hand. The outcome of its machinery lies
in the determination of the much sought after state functions. \\

Unfortunately, for systems that are in steady-states without being in
equilibrium, no such theory exists. As long as the dynamics breaks the
detailed balance condition, steady-state properties will strongly
depend on the details of the microscopic evolution rules, and any
statistical mechanical approach must incorporate those. Yet, a few
guiding principles inherited from equilibrium thermodynamics can be
saved. We shall illustrate this on the example of the granular gas,
that we model as an assembly of hard-spheres interacting only through
inelastic collisions. The irreversible dissipation of energy is
compensated by an energy injection mechanism --two such will be
presented in the sequel-- that maintains the gas in a steady-state,
with time translation invariant properties. Our interest goes to the
total energy $W(t)$ injected by the heating mechanism over an
asymptotically large time window $[0,t]$. This is a space integrated
quantity, as it takes into account the work performed on each
individual particle over $[0,t]$. That $W$ is a space extensive
quantity means that its properties will not be too sensitive to the
microscopic details of the interactions. The key-role of the dynamics
is taken into account in the time extensivity of $W$, but again with the
hope that irrelevant details will be smoothened out through time
integration. These arguments have already been put forward in a series
of experimental and theoretical works bearing on turbulent flows,
convection experiments, or on granular materials themselves. The
belief is that global, {\it i.e.} space and time integrated, observables,
more than local response functions, will allow comparing systems as
far apart as turbulent flows or granular gases whose dynamics
nevertheless share strongly dissipative features, and that the
theorist craves to unite within a common framework.

In the present work we shall focus on the probability distribution
function (pdf) $P(W,t)$ of $W$, as well as on its related generating
function $\hat{P}(\lambda,t)=\langle\ee^{-\lambda W}\rangle$. More
specifically, our efforts will be devoted to determining the
corresponding large deviation functions defined by
\begin{equation}
  \pi(w)=\lim_{t\to\infty}\frac{1}{t}\ln P(W=w\; t,t),\qquad \qquad 
  \mu(\lambda)=\lim_{t\to\infty}\frac{1}{t}\ln \hat{P}(\lambda,t)
\end{equation}
As has been recognized long ago~\cite{ruelle}, $\pi(w)$, or
$\mu(\lambda)$, which are related by a Legendre transform
$\pi(w)=\text{max}_\lambda\{\mu(\lambda)+\lambda w\}$, play the role
of intensive dynamical free energies, solely depending on intensive
variables, like $w(t)=\frac{W(t)}{t}$, the fluctuating
time averaged injected work.\\

It also turns out that in other classes of dynamical systems, the
power injected by the heating mechanism can be related to the entropy
current flowing into the system, whose large deviation function (ldf)
has been shown to possess an important symmetry property: this is the
celebrated fluctuation
theorem~\cite{evanscohenmorriss,gallavotticohen}. While a
crucial hypothesis underlying its demonstration in the aforementioned
works is missing in the granular gas, namely a weak form of time
reversibility, empirical attempts have been made in the past, at an
experimental~\cite{feitosamenon} and a
theoretical~\cite{aumaitrefauvemcnamarapoggi,aumaitrefaragofauvemcnamara} level, to see whether
such a fluctuation relation for $W(t)$ would hold. There is furthermore no straightforward connection between $W$ and entropy production, an otherwise ill-defined concept~\cite{benacoppexdrozviscotrizacvanwijland,puglisiviscotriozacvanwijland} in a granular gas. These issues will
be further discussed in the final section. For now we start by
describing the dynamical evolution rules of our granular gas, along
with two possible energy injection mechanisms. Then we present a
kinetic-theory based approach to determining the related large
deviation functions. The discussion will cast our results within the
 framework of "fluctuation theorems".

\section{Stationary state of a granular gas}
\label{ness}
\subsection{The microscopic dynamics of a granular gas is irreversible}
A standard way of modeling a granular gas is to consider a set of $N$
hard-spheres undergoing inelastic collisions in which a fraction
$(1-\alpha^2)$ of the relative kinetic energy is dissipated away. The
restitution coefficient $\alpha$ lies between $0$ and $1$ (elastic
collisions). Two incoming particles with velocities $\bv_1$ and
$\bv_2$ acquire velocities $\bv_1^*$ and $\bv_2^*$ after having
collided, with
\begin{equation}
  \bv_1^*=\bv_1-\frac{1+\alpha}{2}(\bv_{12}\cdot\bsigma)\bsigma,\qquad
\bv_2^*=\bv_2+\frac{1+\alpha}{2}(\bv_{12}\cdot\bsigma)\bsigma
\end{equation}
The component of the relative velocity $\bv_{12}=\bv_1-\bv_2$ along
the unit vector $\bsigma$ joining the centers of the particles is
reduced by a factor $\alpha$,
$\bv_{12}^*\cdot\bsigma=-\alpha\bv_{12}\cdot\bsigma$. For any given
trajectory in phase space, the time-reversed one is not a physical
trajectory. One can write  an energy balance equation for
the total kinetic energy $E(t)=\sum_i\frac{1}{2}\bv_i^2$, which varies
according to 
\begin{equation} 
\label{energybalance}
\Delta E=E(t)-E(0)=W(t)-D(t)
\end{equation}
where $W(t)$ is the energy injected by an external driving mechanism,
while $D(t)\geq 0$ is the energy irreversibly dissipated through the
inelastic collisions. The average variation rate of $D$ can be estimated as
the collision rate times the energy dissipated through a collision,
\begin{equation}
  \frac{\dd}{\dd t}\langle D\rangle=-\frac{1-\alpha^2}{4\ell}\langle|\bv_{12}\cdot\bsigma|^3\rangle
\end{equation}
where $\ell$ is the mean free path. The mean kinetic energy per
particle provides a typical energy scale, also termed {\it granular
  temperature}, and it is defined as
\begin{equation}
  T_g=\langle\bv_i^2\rangle/d=\beta_g^{-1}
\end{equation}
In the absence of a heating mechanism, dimensional analysis and the assumption of homogeneity, lead to
$T_g$ decaying with time as $T_g(t)\propto t^{-2}$ (Haff's law). We now introduce two relevant heating mechanisms.

\subsection{Free cooling and deterministic isokinetic thermostat}
Each particle $i$, in addition to the inelastic collisions, is
subjected to an energy-injecting viscous friction force $+\gamma {\bf
  v}_i$. However unphysical it may appear at first sight, this
thermostat is actually relevant to the study of the homogeneous
cooling regime~\cite{lutsko,breymontaneromoreno}. During the homogeneous cooling stage, the typical
velocity $\sqrt{T_g}$, and thus the collision frequency, decrease as
$1/t$. Rescaling time with the collision frequency allows to eliminate
time dependence and leads to an effective (nonequilibrium)
steady-state. The latter rescaling is exactly accounted for by an
effective viscous friction force that, instead of dissipating energy,
pumps it into the system. Within the newly rescaled dynamics,
collisions occur at a constant rate. The energy provided by the
thermostat reads
\begin{equation}
W(t)=\sum_i\int_0^t\dd\tau \;\gamma {\bv}_i^2(\tau)
\end{equation}
Note that $W\geq 0$ for all trajectories in phase space. 
\subsection{Heated gas and stochastic thermostat}
A more conventional way of achieving a steady-state is to inject
energy by means of independent random forces acting on each individual
particle:
\begin{equation}
  \frac{\dd \bv_i}{\dd t}={\bf F}_i+\text{collisions},\;\langle F_i^\alpha(t) F_j^\beta(t')\rangle=2\Gamma\delta_{ij}\delta^{\alpha\beta}\delta(t-t') 
\end{equation}
This heating mechanism is easier to handle mathematically than more
realistic thermostats because it leads to a uniform system, by
contrast to boundary drives. It should be noted that there are some
experimental setups that do achieve a uniform
heating~\cite{reisingaleshattuck}. Besides, when the experiment
resorts to vibrating walls, within a large
subvolume far enough from the boundaries, energy and particles
are roughly uniformly distributed~\cite{feitosamenon}, which allows for easier comparison, but this cannot be thought of as resulting from an effective bulk heating mechanism. In the present case, the work
provided by the external heating  reads
\begin{equation}
W(t)=\sum_i\int_0^t\dd\tau \;{\bf F}_i(\tau)\cdot {\bv}_i(\tau)
\end{equation}
Given that $\langle W\rangle/t=2d\Gamma $, the typical energy scale is
set to $T_g=\left(\frac{2d\ell\Gamma\sqrt{\pi}}{(1-\alpha^2)\Omega_d}
\right)^{2/3}$.

\section{Large deviation function for the injected power}
Let $W(t)$ denote the total power injected into the granular gas by
the heating mechanism over the time interval $[0,t]$. We begin with
introducing a phase space density $\rho(\Gamma,W,t)$ that counts the
number of systems in state $\Gamma$ that have accumulated over the
time window $[0,t]$ a total work $W(t)=W$. A generalized Liouville
equation can be written for $\rho$, in which the Liouville operator
${\mathcal L}_W$ can be split into a $W$ conserving part ${\mathcal
  L}_\text{coll}$ and one accounting for changes in $W$ under the
effect of the external injection mechanism ${\mathcal L}_\text{inj}$:
\begin{equation}
  \p_t\rho={\mathcal L}_{W}(\bG,W) \rho={\mathcal
    L}_\text{inj}(\bG,W)\rho+{\mathcal L}_\text{coll}(\bG) \rho
\end{equation}
It is a convenient detour to go first to the Laplace transform of
$\rho$, $\hat{\rho}(\Gamma,\lambda,t)=\int\dd W\ee^{-\lambda
  W}\rho(\Gamma,W,t)$, then rewrite the Liouville equation in terms of
$\hat{\rho}$,
\begin{equation}\label{notLiou}
  \p_t\hat{\rho}={\mathcal{L}_{W}}(\bG,
  \lambda)\hat{\rho}={\hcL}_\text{inj}(\bG,
  \lambda)\hat{\rho}+{\hcL}_\text{coll}(\bG) \hat{\rho}
\end{equation}
The largest eigenvalue $\mu(\lambda)$ of ${\mathcal L}_W(\lambda)$ thus
governs the asymptotic behavior of $\hat{\rho}$,
\begin{equation}
  \hat{\rho}(\Gamma,\lambda,t)\simeq C(\lambda) \ee^{\mu(\lambda)t}\tilde{\rho}(\Gamma,\lambda)
\end{equation}
where $\tilde{\rho}(\Gamma,\lambda)$ is the (right) eigenvector of
${\hcL_{W}}(\bG,\lambda)$ associated to $\mu(\lambda)$, and
$C(\lambda)$ is the projection of the initial state on this
eigenvector. We have chosen to normalize $\tilde{\rho}$ to unity,
which, given that it has a definite sign, endows it with the meaning
of a probability distribution (note however that (\ref{notLiou}) is a
not a Liouville equation since it does not conserve probability). We
shall provide further insight on this later. In order to determine
$\mu$ and $\tilde{\rho}$, we project the eigenvalue equation onto the
one particle subspace. We shall focus on this
isokinetic thermostat for which the results presented here are new. We
arrive at
\begin{equation}\label{eveq}
  \mu(\lambda)\tilde{f}^{(1)}(\bv_1,\lambda)=-\gamma\p_{\bv_1}\cdot\left(\bv_1\tilde{f}^{(1)}(\bv_1,\lambda)\right)-\lambda\gamma v_1^2\tilde{f}^{(1)}(\bv_1,\lambda)+\text{ coll .}
\end{equation}
where in the rhs of (\ref{eveq}) the loose notation ``coll'' is a
shorthand for the full collision operator acting on the two point
function $\tilde{f}^{(2)}$. In order to extract physical information
from this equation, we resort to the molecular chaos hypothesis, which
turns the two-body interaction term into
\begin{equation}
  \text{ coll .}=\frac{1}{\ell}\int_{\bv_{12} \cdot \bsigma >0}
  \dd \bv_2\,\dd\bsigma \,(\bv_{12}\cdot \bsigma)\,\left(\frac{1}{\alpha^2}
    f^{(1)}(\bv_1^{**},\lambda)f^{(1)}(\bv_2^{**},\lambda)-
    f^{(1)}(\bv_1,\lambda)f^{(1)}(\bv_2,\lambda)\right)
\end{equation}
Not surprisingly, one recovers the steady-state velocity pdf equation
at $\lambda=0$.  It is interesting that the process encoded in
(\ref{eveq}) can be read off as the original granular gas dynamics
in which additional particles are created (or destroyed, according to
whether $\lambda<0$ or $>0$) at a velocity dependent rate $\lambda
v^2$.  The eigenvalue $\mu(\lambda)$ is then interpreted as the
population growth rate. This remark was numerically exploited, for
somewhat different systems, by Giardin\`a, Kurchan and Peliti
\cite{gkp}. The splitting of $f^{(2)}$ as a product of independent
one particle distributions is indeed a molecular chaos hypothesis for
the system with the non particle conserving fictitious dynamics. The
Boltzmann equation toolbox offers many ways to arrive at an expression
for $\mu(\lambda)$. \\

Let us start with $\lambda\sim 0$, that is in a regime for which $W$
lies in the vicinity of $\langle W\rangle$. Strictly at $\lambda=0$,
that is when $\tilde{f}^{(1)}(\bv,0)$ yields the steady-state velocity
pdf, Sonine expansions have successfully been used to characterize
deviations from the Maxwell distribution \cite{vannoije}. These
expansions work all the better as space dimension $d$ is high, given
that the phase space contraction due to the inelastic collisions
occurs only along one --among $d$-- space direction.  The coefficients
of the expansion turn out to be functions of the reduced variable
$(1-\alpha^2)d^{-1}$, thus making explicit that a large $d$ expansion
is equivalent to a quasi-elastic limit, hence the success of expanding
around a Gaussian.  What cannot be guessed, however, is that in
practice $d=2$ or $d=3$ already turn out high enough dimensions for the
Sonine expansions to provide quantitatively reliable results.  This
provides the necessary motivation for attempting a Sonine expansion at
$\lambda \sim 0$.  However, further analysis, in the same vein as that
carried out by Van Noije and Ernst \cite{vannoije}, shows that in the
large velocity limit (and at $\lambda >0$), the one point function
$\tilde{f}^{(1)}(\bv,\lambda)$ must be a Gaussian. The simplest
approximation is thus to project $\tilde{f}^{(1)}(\bv,\lambda)$ onto a
Gaussian with a $\lambda$-dependent variance. From (\ref{eveq}) one
can easily see that
\begin{equation}\label{linkmuv} \mu(\lambda)/N=-\gamma \lambda\langle
  {\bf
    v}^2\rangle_\lambda 
\end{equation} 
where $\langle \ldots\rangle_\lambda$ denotes an average performed
with respect to the weight given by $\tilde{f}^{(1)}$. This equation
relates the variance of $\tilde{f}^{(1)}$ to $\mu$ in a particularly
simple way. Thus, working in terms of rescaled quantities, with the
granular temperature precisely given by
\begin{equation} T_g=\left( \frac{2 d \ell \gamma \sqrt{\pi}}{(1-\alpha^2)
\Omega_d} \right)^2\,\,.  \end{equation} and performing the following
replacements \begin{equation} \mu:=\mu/(N\gamma),\;\;\lambda:=\lambda T_g
d/2 \end{equation} one has, for the dimensionless quantities:
\begin{equation} \mu(\lambda)= -\frac{ 1 + 2\,\lambda - {\sqrt{1 +
4\,\lambda}}  }{2\, {\lambda}} \end{equation} The asymptotic behavior is
given by: \begin{equation} \mu(\lambda) \stackrel{\lambda \to -1/4}{\sim}
1-2 \sqrt{1+4 \lambda} +{\mathcal O}\left(\lambda+\frac{1}{4}\right)\,\,,\qquad \qquad\\
\mu(\lambda) \stackrel{\lambda \to \infty}{\sim} -1 +
\frac{1}{\sqrt{\lambda}} + {\mathcal O}(\frac{1}{\lambda})
\end{equation} 
Taking (\ref{linkmuv}) into account one sees that the
typical temperature scale as given by $\tilde{f}^{(1)}$ is given by
$T(\lambda)\sim T_g \left|\frac{\mu(\lambda)}{ 2\lambda}\right|$. That
$T(\lambda\to+\infty)\to 0$ means that values of $W$ much lower than the
average $\langle W\rangle$ are produced during trajectories over which
typical velocities are small, which is of course  no surprise.  By
contrast, given that $T(\lambda\to-1/4)=2T_g$, the larger than average $W$ trajectories arise from realizations
in which the effective kinetic temperature is at a value twice the
stationary state temperature $T_g$.\\

It is possible to express the large deviation function of $w=W/t$ by a
simple Legendre transform, 
\begin{equation} \pi(w) \stackrel{w \to
    0^+}{\sim} -1 + \frac{3}{2^{2/3}} w^{1/3} + {\mathcal
    O}(w^{2/3})\,\,, \qquad \qquad \pi(w) \stackrel{w \to
    \infty}{\sim} -\frac{w}{4} + {\mathcal O}(\sqrt{w})
\end{equation} 
but the validity of the $w\to 0^+$ expression may be hindered by subtle
effects that we shall discuss in the next section.

\begin{figure}[t] 
\begin{minipage}[t]{.46\linewidth}
\includegraphics[clip=true,width=1 \textwidth]{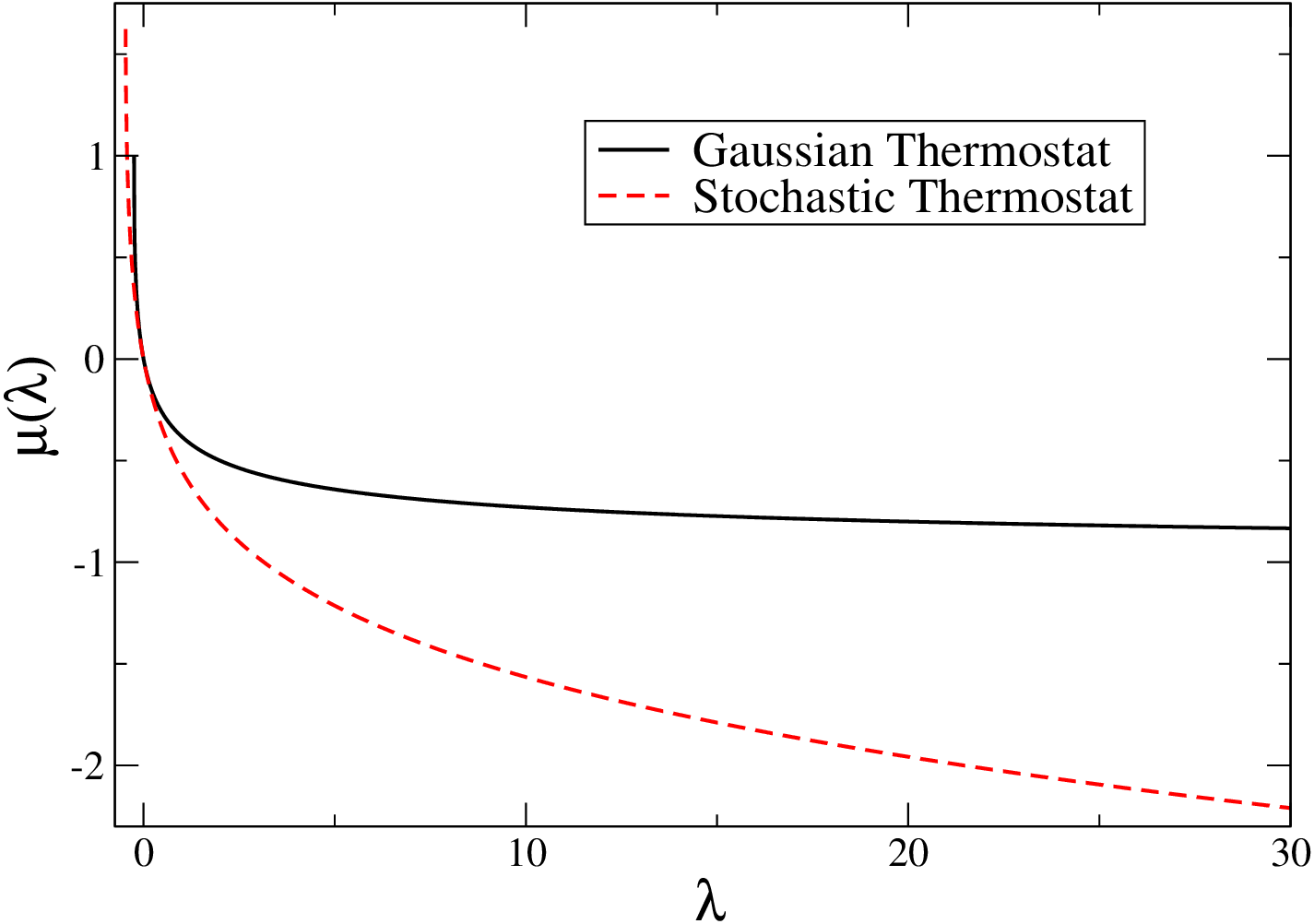}
\caption{\label{fig:mu_gauss} Plot of $\mu(\lambda)$.} 
\end{minipage}
\hfill 
\begin{minipage}[t]{.46\linewidth}
\includegraphics[clip=true,width=1 \textwidth]{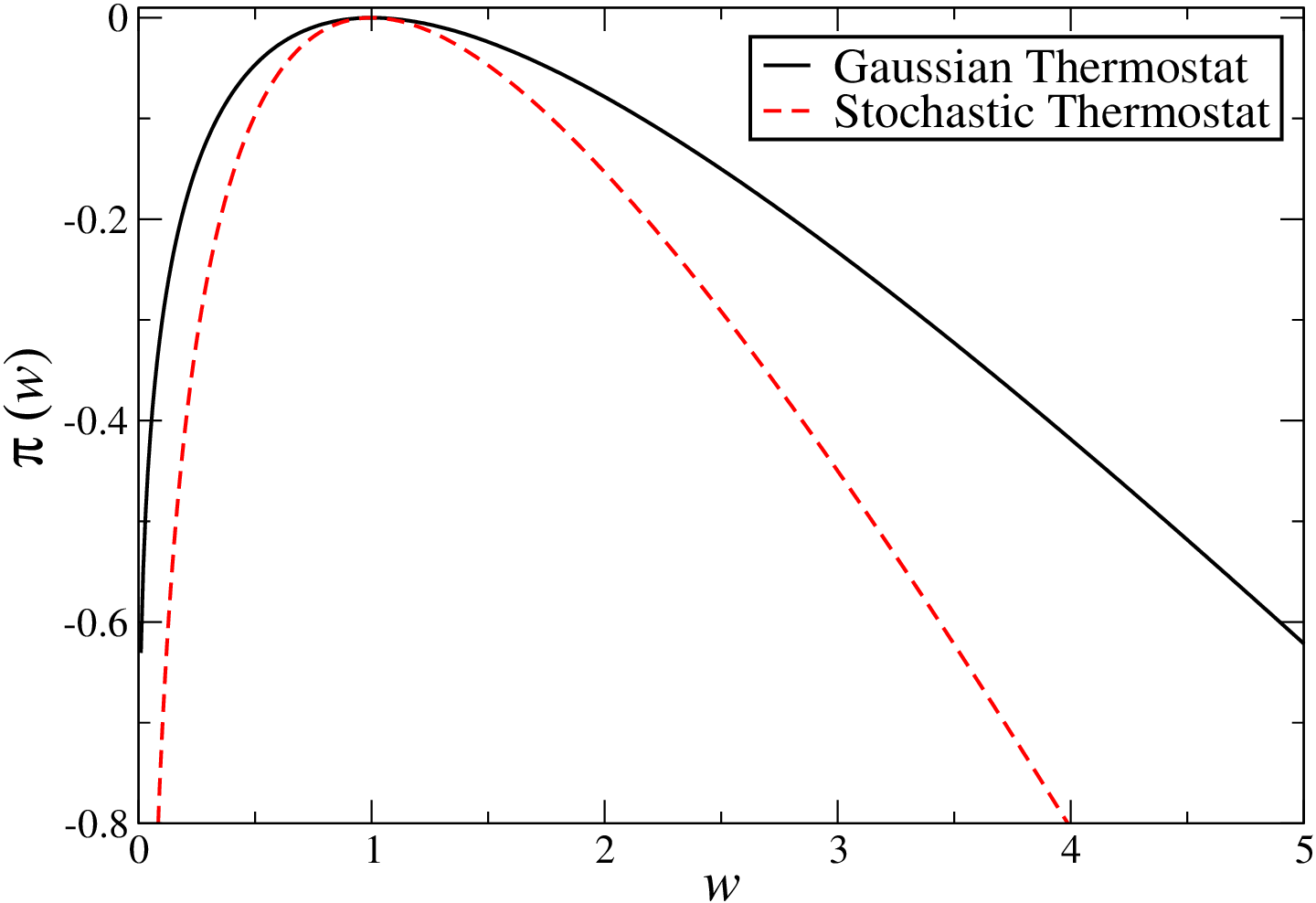} 
\caption{Plot
of $\pi(w)$.} 
\end{minipage} 
\end{figure}

We refer the reader to
\cite{viscopuglisibarrattrizacvanwijland1,viscopuglisibarrattrizacvanwijland2}
for details about the stochastic thermostat for which we obtain, in
terms of dimensionless variables (for which $\langle w\rangle=1$), 
\begin{equation}\label{scales}
  \mu:=\mu\frac{T_g}{d\Gamma N}\,\,, \qquad \qquad
\lambda:=\lambda T_g 
\end{equation}
the following expression
\begin{equation}
\mu(\lambda)=-\lambda+\frac{1}{2}\frac{T(\lambda)}{T_g}\lambda^2
\end{equation} 
where $\sqrt{T(\lambda)}$ sets the typical velocity scale
for trajectories characterized by $\lambda$. At large values of $\lambda$,
corresponding to values of $W$ small with respect to $\langle
W\rangle$, we expect that $T(\lambda)\ll T_g$ and indeed 
$\frac{T(\lambda)}{T_g}\simeq \frac{2}{\lambda}$ as
$\lambda\to+\infty$. 
This can be further refined to
obtain the behavior of $\mu(\lambda)$ at $\lambda\to+\infty$,
\begin{equation} 
  \mu(\lambda)\stackrel{\lambda\to+\infty}{\simeq}-\lambda^{1/4} 
\end{equation} 
Besides, $\mu(\lambda)$ possesses a cut in the $\lambda$ plane at
$\lambda_c=-3/2^{8/3}$, such that, as $\lambda\to\lambda_c^+$,

\begin{equation} 
  \mu(\lambda)=3/
  2^{2/3}-3^{3/2}2^{1/6}\sqrt{\lambda-\lambda_c}+{\mathcal
    O}(\lambda-\lambda_c) 
\end{equation} 
The presence of this cut is responsible for the exponential decay of
the pdf of $W$ at large values of $W$. However, the non-analytic
behavior at $\lambda\to+\infty$ leads to a non-analytic behavior as
$w=W/t\to 0^+$,
\begin{equation}
  \pi(w\to 0^+)\sim-w^{-1/3}\,\,,\qquad 
\pi(w\to 1) \simeq -(w-1)^2/2\,\,,\qquad
\pi(w\to\infty)\sim -w
\end{equation}

\section{Dissipated energy}
\subsection{Cumulant generating function}
So far our interest has been focused on the distribution of the energy
injected into the system to keep it in a nonequilibrium steady state.
As already mentioned in section \ref{ness}, the dissipated energy is
intimately related to the injected power trough Eq
(\ref{energybalance}), which expresses the energy balance of the
system, and which we rewrite here for a practical purpose:
\begin{equation}
\label{balance}
\Delta E(t) = W(t) - D(t)
\end{equation}
Since we are interested in the large time behavior of both $W(t)$ and
$D(t)$, it is worthy to note that while $W(t)$ and $D(t)$ are of order
$t$, the boundary term $\Delta E(t)$ is of order one. This simple
remark would intuitively, and na\"{\i}vely, lead to the conclusion that
at large time $W(t)$ and $D(t)$ are distributed in the same way, and
hence share the same large deviation function. Before entering  the
details of the reasons for which $W$ and $D$ do not necessarily share the
same large deviation function, we will first show that  they actually 
share the same cumulant generating function.

The dissipated energy $D(t)$ increases at each collision by an amount
equal to the difference of the energy of the colliding pair just
before (at $t=t^-$) and immediately after ($t=t^+$) the collision:
\begin{equation}
D(t^+) = D(t^-) + \frac{1-\alpha^2}{4} (\bv_{ij} \cdot \bsigma)^2 
\end{equation}
If we now define the joint probability $\rho_D(\bG, D,t)$ of having a
microscopic configuration $\bG$ and a given value of $D$ at time $t$, 
it is possible to write a Liouville equation for $\rho_D$:
\begin{equation}\label{evolD}
\frac{\p}{\p t} \rho_D(\bG,D,t) = \cL_{D}(\bG, D) \rho_D(\bG,D,t)= \cL_{\text{inj}}(\bG) \rho_D + \cL_{\text{diss}}(\bG, D) \rho_D  
\end{equation}
where  $\cL_{\text{inj}}$ describes some energy injection mechanism. As for the collision operator, its explicit expression reads 
\begin{equation}
\cL_{\text{diss}}=\sigma^{d-1} \sum_{i<j}^N \int_{\bv_{ij} \cdot \bs >0} \dd \bs (\bv_{ij} \cdot
\bs) \left( \frac{1}{\alpha^2} \delta(\br_{ij}- \bsig) {\tilde b}^{**}_{ij} -
  \delta(\br_{ij}+ \bsig) \right)
\end{equation}
where $\sigma$ is the particle diameter and where
\begin{equation}
{\tilde b}^{**}_{ij} g \left(\bv_1,\dots,\bv_i,\dots,\bv_j,\dots,\bv_N,
D\right)=
g\left(\bv_1,\dots,\bv^{**}_i,\dots,\bv^{**}_j,\dots,\bv_N,D-
\frac{(\bv_{ij} \cdot \bs)^2}{4} 
\left( \frac{1}{\alpha^2} -1 \right)\right)
\end{equation}
We now translate (\ref{evolD}) in terms of  the Laplace transform $\hr_D(\bG,\lambda,t) = \int \dd D
\ee^{-\lambda D}\rho_D(\bG,D,t)$, which evolves according to
\begin{equation}
\frac{\p}{\p t} \hr_D(\bG,\lambda, t)=\cL_{\text{inj}}(\bG) \hr_D + \hcL_{\text{diss}}(\bG, \lambda) \hr_D
\end{equation}
with 
\begin{equation}
\hcL_{\text{diss}}(\bG, \lambda)=\sigma^{d-1} \sum_{i<j}^N \int_{\bv_{ij} \cdot \bs >0} \dd \bs (\bv_{ij} \cdot
\bs) \bigg( \frac{1}{\alpha^2} \delta(\br_{ij}- \bsig) 
\ee^{-\lambda \frac{(\bv_{ij} \cdot \bs)^2}{4} 
\left( \frac{1}{\alpha^2} -1 \right)} b^{**}_{ij} -
  \delta(\br_{ij}+ \bsig) \bigg)
\end{equation}
and
\begin{equation}
{\tilde b}^{**}_{ij} g \left(\bv_1,\dots,\bv_i,\dots,\bv_j,\dots,\bv_N \right) =
g\left(\bv_1,\dots,\bv^{**}_i,\dots,\bv^{**}_j,\dots,\bv_N \right)
\end{equation}
A straightforward calculation shows that
\begin{equation}\label{simildiss}
\hcL_{\text{diss}}(\bG, \lambda)=\ee^{-\lambda\Delta E}\hcL_{\text{diss}}(\bG, 0)\ee^{+\lambda\Delta E}=\ee^{-\lambda\Delta E}\hcL_{\text{diss}}(\bG)\ee^{+\lambda\Delta E}
\end{equation}
Assuming, for definiteness, a deterministic energy injection mechanism in which each particle is subjected to a force ${\bf F}_i$, we must have
\begin{equation}
\hcL_{\text{inj}}(\bG, \lambda)=-\lambda\sum_i{\bf F}_i\cdot{\bf v}_i-\sum_i\p_{{\bf v}_i}\cdot({\bf F}_i\;)
\end{equation}
Hence, conversely to (\ref{simildiss}), we may verify that,
\begin{equation}
\hcL_{\text{inj}}(\bG, \lambda)=\ee^{+\lambda\Delta E}\hcL_{\text{inj}}(\bG, 0)\ee^{-\lambda\Delta E}=\ee^{+\lambda\Delta E}\hcL_{\text{inj}}(\bG)\ee^{-\lambda\Delta E}
\end{equation}
Using that $ \hcL_{W}(\bG,\lambda)=\hcL_{\text{inj}}(\bG, \lambda)+\hcL_{\text{diss}}(\bG, 0)$ and that $\hcL_{D}(\bG,\lambda)=\hcL_{\text{inj}}(\bG, 0)+\hcL_{\text{diss}}(\bG, \lambda)$ we arrive at the following key identity
\begin{equation}\label{simil}
  \hcL_{W}(\bG,\lambda)= \ee^{-\lambda \Delta E} 
  \hcL_{D}(\bG, \lambda) \ee^{\lambda \Delta E}
\end{equation}
That  $ \hcL_{D}(\bG,\lambda)$ and $ \hcL_{W}(\bG,\lambda)$ are related through the similarity transformation (\ref{simil})  establishes that these operators have exactly the same eigenvalues, and that the
corresponding eigenvectors differ only by a factor $\ee^{\lambda
  \Delta E}$. This allows us to conclude that $W$ and $D$ have the same
cumulant generating function. Thus, if furthermore the largest
eigenvalue of $\hcL_{D}(\lambda)$ is related to the large deviation
function for the dissipated energy through a Legendre transform, the
above analysis shows that the large deviation function of the injected
and dissipated energy are indeed equal.

\subsection{Influence of time boundary contributions} 

It has already been stressed~\cite{farago} that it is possible that
$\mu(\lambda)$ and $\pi(w)$ may not be related by a Legendre
transform. As already discussed in the previous subsection, one
expects, in writing the energy balance equation (\ref{balance}), that,
given that $\Delta E$ is not extensive in time, $W$ and $D$ should
have the same large deviation functions (expressed in terms of $w=W/t$
or $\delta=D/t$). Since they share the same
cumulant generating function $\mu(\lambda)$, this also means that
within a finite interval around their average value, $w$ and $d$ do
share the same large deviation function. Nevertheless, beyond a finite
value of $w$ or $\delta$, their ldf may become distinct; this may
occur~\cite{farago,vanzoncohen,bonettogallavottigiulianizamponi,visco}
if $\Delta E$ is distributed exponentially or slower than
exponentially. The technical reason of this phenomenon is simply due
to a problem of Laplace transform inversion. In fact, we assumed that
the ldf $\pi$ is related to the eigenvalue $\mu$ trough a Legendre
transform. This result is obtained carrying out the Laplace transform
inversion trough a saddle point expansion (in the $t \to \infty$
limit). In practice this last step is valid only for values of
$\lambda$ for which it is possible to define a path (in the
$\lambda$-complex plane) including a straight line parallel to the
$\lambda$-imaginary axis. The problem arises hence when there are some
cuts which are not included in the expression of $\mu(\lambda)$, but in
some other \emph{subleading} term. Unfortunately, as far as we know, there is no
general method to know {\it a priori} if and where the two large
deviation functions may differ. The only argument in this direction
involves the probability distribution of the boundary term: as already
mentioned above, if its tails decrease exponentially, or slower, it is
possible that such problem may occur.
This is the scenario in the cases at hand. One can show
that $\Delta E$ has tails decreasing as $\exp(-N^{1/2}(\beta_g \Delta
E)^{1/2})$ and $\exp(-N^{1/4}(\beta_g \Delta E)^{3/4})$ for the
Gaussian thermostat and for the stochastic thermostat, respectively (an analysis of the typical values of $\Delta E$ is presented in \cite{viscopuglisibarratvanwijlandtrizac}).
Note also that these tails decrease exponentially in the number of
degrees of freedom, which leaves room, in principle, even in the
thermodynamic limit, for $W$ and $D$ having ldf differing
significantly beyond a finite threshold value.

\section{A critical discussion of numerical results} 

\begin{figure}[t] 
\begin{minipage}[t]{.46\linewidth}
\includegraphics[clip=true,width=1 \textwidth]{gc_tau.eps}
\caption{\label{fig:gc} Plot of $\pi(w,t)-\pi(-w,t)$ for $t=1,2$ and
  $3$ mean free times (mft). The inset show the probability density
  function of $w(t)$ for the same times.} 
\end{minipage}
\hfill 
\begin{minipage}[t]{.46\linewidth}
\includegraphics[clip=true,width=1 \textwidth]{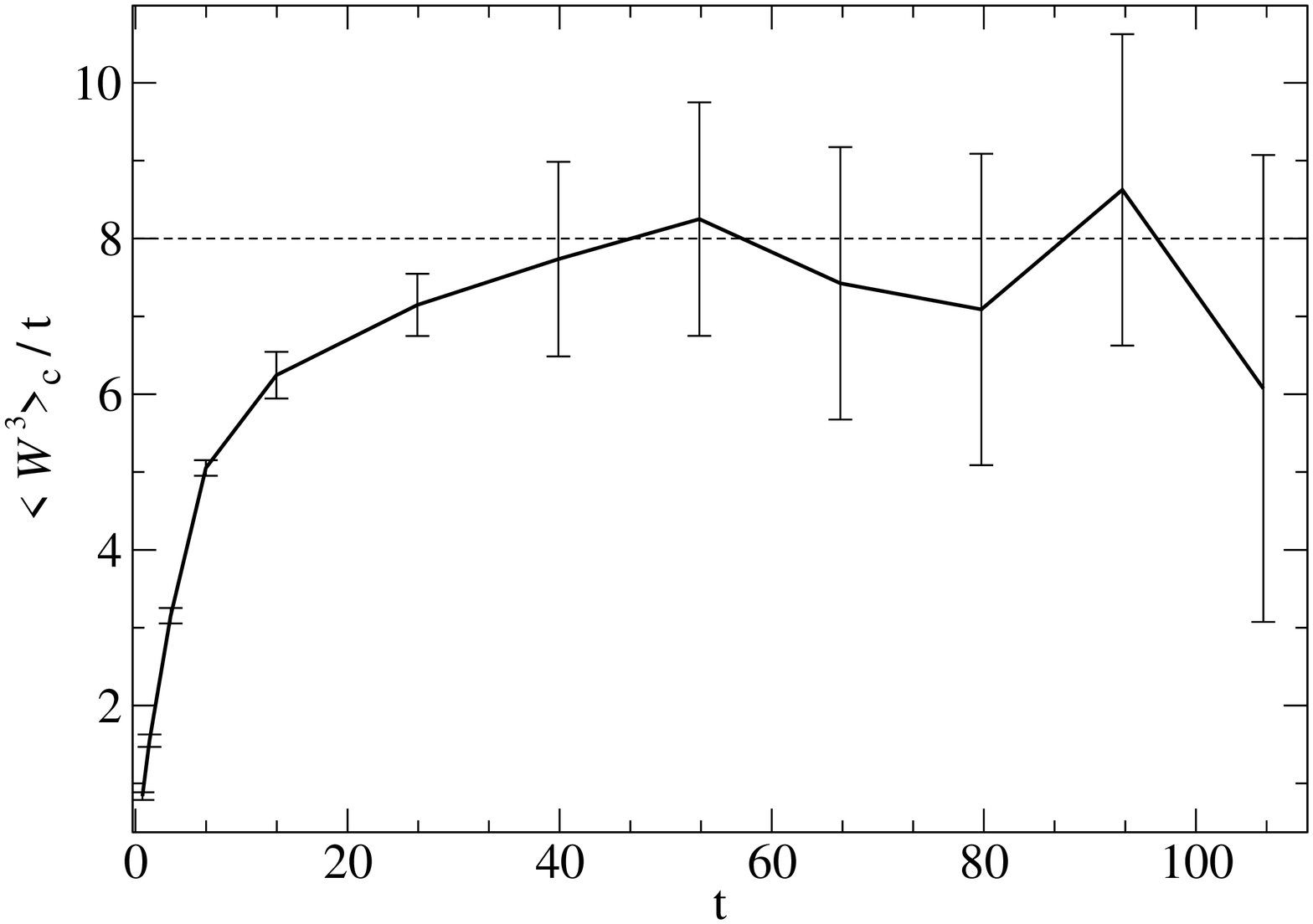} 
\caption{\label{fig:cumulant}Plot
of the third cumulant of $W$ divided by time. The asymptotic regime is reached after fifty mft.} 
\end{minipage} 
\end{figure}

In this section we would like to present simulations of the granular
gas heated with the stochastic thermostat. The total work $W(t)$
provided by the random forces over $[0,t]$ is of course, on average, a
positive quantity, but there exist phase space trajectories which will
yield a negative $W$ (this cannot occur for the deterministic
thermostat where $W(t)\geq 0$).  It is therefore tempting to plot the
quantity $\pi(w)-\pi(-w)$ as a function of $w$. It is
well-known~\cite{evanscohenmorriss,gallavotticohen} that for a
well-defined class of thermostatted systems the power $W$ injected by
the thermostat verifies the celebrated fluctuation theorem
$\pi(w)-\pi(-w)=\beta w$, where $\beta$ is an appropriate inverse
energy scale. However, one of the key hypotheses underlying the latter
fluctuation relation is that the dynamics features a weak form of time
reversibility, for which each trajectory possesses an time reversed
partner (however unlikely). In our granular gas, dissipative
collisions hamper time reversed trajectories to be physically
acceptable trajectories at all. It should therefore come as no
surprise that no specific fluctuation relation emerges in our case. Of
course, this is fully confirmed by the explicit calculations presented
in the previous section, that show no specific symmetry property of
the injected power ldf.  In spite of this, it may be instructive to
plot $w\mapsto\frac{1}{t}\ln \frac{P(w t,t)}{P(-wt,t)}$ at large times
(in the infinite time limit, this is exactly $\pi(w)-\pi(-w)$), as has
been done in a vibrated granular gas experiment~\cite{feitosamenon}.
This is shown in figure \ref{fig:gc}, and the result is rather intriguing:
one actually observes a straight line with slope $\beta_g$! This
deserves to be explained, given that our analytic results establish
the absence of such a relation, even for values of $w$ not too far
from its average. The first remark is that a numerical simulation is
always carried out at finite times, and thus one is measuring
$\pi(w,t)=\frac{1}{t}\ln P(w t,t)$ rather than its $t\to\infty$ limit
$\pi(w)$. Hence the first question is: has the simulation reached the
infinite time regime? The answer to that question is tricky. Though
$\pi(w,t)$ notably deviates from a quadratic form, which would
correspond to $P(W,t)$ being Gaussian, this is no proof that the
asymptotic regime has been reached. For that matter it is instructive
to investigate the behavior of the third cumulant $\langle
W^3\rangle/t$ as a function of time. This is presented in figure
\ref{fig:cumulant}. The conclusion is that the third cumulant reaches its
asymptotic value, consistent with the analytic expression, at times a few tens as large as those for which it is possible to measure negative values of $w=W/t$, as presented in figure \ref{fig:gc}. A
similar plot of the fourth cumulant would signal that the asymptotic
regime has not been reached over the chosen time window. What one
actually observes is simply the leftover of a short time quadratic
behavior for $\pi(w)$. This is also consistent with the quadratic
approximation for $\mu(\lambda)$, which, if taken for granted as the
whole function, would indeed imply $\pi(w)-\pi(-w)=\beta_g w$ (after restoring the appropriate physical scales as in (\ref{scales})). The
lesson to be drawn from this discussion is that it seems to be an
optimistic endeavor to investigate solely on numerical or experimental
grounds the validity of the fluctuation relation. Without criteria for
the time scale at which the asymptotic regime is entered, the analysis is at risk of remaining
confined within short time
effects~\cite{feitosamenon,puglisiviscobarrattrizacvanwijland}.

\section{Conclusion} 
We have presented an analytic calculation for the large deviation
function of an $N$-body dynamical system with strong dissipative
interactions. By contrast to existing approaches, based on a
stochastic modeling, the noise source is completely contained within
the microscopic formulation, which avoids the arbitrariness inherent
to choosing a particular type of Markov dynamics. Our calculation
required that we extend the standard methods of kinetic theory, so as
to grasp the infinite hierarchy of correlation functions encoded in a
temporal large deviation function. We have shown that theoretical input was required to properly analyze numerical data when large deviations are measured. Our study also allows us to infer that
the power injected into a granular gas to maintain a steady-state has
both generic features and ones that are sensitive to the details of
the mechanisms at work. At large injected powers, the ldf $\pi(w)$
behaves linearly with $w=W/t$, meaning that the power distribution
decays exponentially. A less robust feature, however, is the $w\to 0$
behavior, which proves extremely sensitive to the details of the
heating mechanism. The present work calls for further investigation in other dissipative systems, such as turbulent flows.



\section*{Acknowledgements}
The authors acknowledge useful discussions with F. Zamponi along with the financial support of the French Ministry of Education through an ANR grant JCJC-CHEF.


\begin{thebibliography}{00}

\bibitem{ruelle} D. Ruelle, {\it Thermodynamic formalism}, Addison-Wesley, (Reading, Mass., 1978).

\bibitem{evanscohenmorriss} D.J. Evans, E.G.D. Cohen and G.P. Morriss,
  Phys. Rev. Lett. 71, 2401 (1993) and ibid. 3616 (1993).

\bibitem{gallavotticohen} G. Gallavotti and E.G.D. Cohen, Phys. Rev.
  Lett. {\bf 74}, 2694 (1995).

 \bibitem{feitosamenon} K. Feitosa and N. Menon, Phys. Rev. Lett. {\bf
    92}, 164301 (2004).
    
\bibitem{aumaitrefauvemcnamarapoggi} S. Auma\^{\i}tre, S. Fauve, S. Mc
  Namara, P. Poggi, Euro. Phys. J. B {\bf 19}, 449 (2001). 
  
\bibitem{aumaitrefaragofauvemcnamara} S. Auma\^{\i}tre, J. Farago, S.
  Fauve and S. Mc Namara, Eur. Phys. J. B {\bf 42}, 255 (2004).  
  
 \bibitem{benacoppexdrozviscotrizacvanwijland} I. Bena, F. Coppex, M.
  Droz, P. Visco, E. Trizac, F. van Wijland, Physica A {\bf 37}, 179
  (2006).
  
 \bibitem{puglisiviscotriozacvanwijland} A. Puglisi, P. Visco, E.
  Trizac and F. van Wijland, Phys. Rev. E {\bf 73}, 021301 (2006).
 
 \bibitem{lutsko}
 J.F. Lutsko, Phys. Rev. E {\bf 63}, 061211 (2001).
 
 \bibitem{breymontaneromoreno}
 J. J. Brey, M. J. Ruiz-Montero and F. Moreno, Phys. Rev. E {\bf 69}, 051303 (2004).
  
\bibitem{reisingaleshattuck} P.M. Reis, R.A. Ingale and M. D. Shattuck, Phys. Rev. Lett. {\bf 96}, 258001 (2006).

\bibitem{gkp} C. Giardin\`a, J. Kurchan and L. Peliti,
Phys. Rev. Lett. {\bf 96}, 120603 (2006). 

\bibitem{vannoije} T. P. C. van Noije and M. H. Ernst, Gran. Matt.
  {\bf 1}, 57 (1998).
 
\bibitem{viscopuglisibarrattrizacvanwijland1} P. Visco, A. Puglisi, A.
  Barrat, E. Trizac and F. van Wijland, Europhys. Lett. {\bf 72}, 55
  (2005).

\bibitem{viscopuglisibarrattrizacvanwijland2} P. Visco, A. Puglisi, A.
  Barrat, E. Trizac and F. van Wijland,  J. Stat. Phys. {\bf 125}, 529 (2006).
    
\bibitem{farago} J. Farago, J. Stat. Phys. {\bf 107}, 781 (2002).

\bibitem{vanzoncohen} R. van Zon and E.G.D. Cohen, Phys. Rev. Lett.
  {\bf 91}, 110601 (2003).

\bibitem{bonettogallavottigiulianizamponi} F.Bonetto, G.Gallavotti,
  A.Giuliani and F.Zamponi, J.Stat.Phys. {\bf 123}, 39 (2006).

\bibitem{visco}
P.  Visco, J. Stat. Mech., P06006 (2006).

\bibitem{viscopuglisibarratvanwijlandtrizac} P. Visco, A. Puglisi, A.
  Barrat, F. van Wijland, E. Trizac, Eur. Phys. J. B {\bf 51}, 377
  (2006).

 \bibitem{puglisiviscobarrattrizacvanwijland} A. Puglisi, P. Visco, A.
  Barrat, E. Trizac and F. van Wijland, Phys. Rev. Lett. {\bf 95},
  110202 (2005). 
  
  
  
  
  









  

  


  

 



\end{thebibliography}
\end{document}